\documentclass[amsmath,amssymb,prl,hyperlink,twocolumn]{revtex4}

\usepackage{qcircuit}
\usepackage[dvips]{graphicx}
\usepackage{amsmath,amssymb,amsthm,mathrsfs,amsfonts,dsfont}
\usepackage{subfigure, epsfig}
\usepackage{braket}
\usepackage{bm}
\usepackage{enumerate}
\usepackage{color}

\begin{document}
\title{Demonstration of Adiabatic Variational Quantum Computing with a \\ Superconducting Quantum Coprocessor}

\date{\today}

\author{Ming-Cheng Chen$^{1,2,*}$}
\author{Ming Gong$^{1,2,*}$}
\author{Xiao-Si Xu$^{3,*}$}
\author{Xiao Yuan$^{3}$}
\author{Jian-Wen Wang$^{1,2}$}
\author{Can Wang$^{1,2}$}
\author{Chong Ying$^{1,2}$}

\author{Jin Lin$^{1,2}$}
\author{Yu Xu$^{1,2}$}

\author{Yulin Wu$^{1,2}$}
\author{Shiyu Wang,$^{1,2}$}

\author{Hui Deng$^{1,2}$}
\author{Futian Liang$^{1,2}$}
\author{Cheng-Zhi Peng$^{1,2}$}

\author{Simon C. Benjamin$^{3}$}
\author{Xiaobo Zhu$^{1,2}$} %\email{xbzhu16@ustc.edu.cn}
\author{Chao-Yang Lu$^{1,2}$}  %\email{cylu@ustc.edu.cn}
\author{Jian-Wei Pan$^{1,2}$ \vspace{0.2cm} } 

\affiliation{$^{1}$Shanghai Branch, National Laboratory for Physical Sciences at Microscale and Department of Modern Physics, University of Science and Technology of China, Shanghai 201315}
\affiliation{$^{2}$CAS Center for Excellence in Quantum Information and Quantum Physics, University of Science and Technology of China, Hefei, Anhui 230026, China}
\affiliation{$^{3}$Department of Materials, University of Oxford, Parks Road, Oxford OX1 3PH, United Kingdom}

\maketitle

\textbf{Adiabatic quantum computing enables the preparation of many-body ground states~\cite{RevModPhys.90.015002}. This is key for applications in chemistry~\cite{aspuru2005simulated,helgaker2014molecular,2018arXiv180810402M,cao2018quantum}, materials science~\cite{lanyon2011universal,ma2011quantum,PhysRevX.8.011044}, and beyond~\cite{ulungu1994multi,farhi2001quantum,pop2007exact}. Realisation poses major experimental challenges: Direct analog implementation requires complex Hamiltonian engineering, while the digitised version needs deep quantum gate circuits. 
%Recently the variational quantum-classical hybrid algorithm has emerged as a means to use low-depth parameterised circuits to approximate many-body ground states. However this route suffers from the hard problem of global multi-parameter optimisation. 
To bypass these obstacles, we suggest an adiabatic variational hybrid algorithm~\cite{Li2017,2018arXiv181208767Y}, which employs short quantum circuits and provides a systematic quantum adiabatic optimisation of the circuit parameters. The quantum adiabatic theorem promises not only the ground state but also that the excited eigenstates can be found. 
We report the first experimental demonstration that many-body eigenstates can be efficiently prepared by an adiabatic variational algorithm assisted with a multi-qubit superconducting coprocessor. We track the real-time evolution of the ground and exited states of transverse-field Ising spins with a fidelity up that can reach about 99\%. }

%The efficient computation of low-lying eigenstates of a many-body quantum Hamiltonian is crucial to the study of quantum chemistry~\cite{aspuru2005simulated,helgaker2014molecular,2018arXiv180810402M,cao2018quantum}, quantum materials~\cite{lanyon2011universal,ma2011quantum,PhysRevX.8.011044} and the applications of classical combinatorial optimization~\cite{ulungu1994multi,farhi2001quantum,pop2007exact}. %Computational complexity theory shows that the ground-state problem of a general many-body Hamiltonian belongs to the quantum-Merlin-Arthur (QMA) class~\cite{kempe2006complexity,Oliveira:2008:CQS:2016985.2016987,schuch2009computational,aaronson2009computational}, which contains the set of classical NP-hard problems. 
It is believed that there are no classical algorithms for efficiently solving the general quantum ground state problems due to the notorious sign-problem~\cite{PhysRevB.41.9301}. In contrast, quantum computing avoids the sign-problem by directly operating with quantum states and thus may provide a profound speedup~\cite{Lloyd1073,Abrams97}. 
Adiabatic state preparation is a natural approach for quantum ground-state problems~\cite{farhi2001quantum,aspuru2005simulated}. Starting from the ground state  of a simple initial Hamiltonian $H_0$, such as $\ket{+}^{\oplus N}$ of $H_0=-\sum^N_{i=1}\sigma^{i}_x$, we would evolve to a complex target Hamiltonian $H_T$. 
The quantum adiabatic theorem~\cite{Born1928} guarantees that if the change is sufficiently slow, the system will stay at its instantaneous eigenstate and ultimately reach the ground state of $H_T$. In certain cases it may be possible to realise the evolving Hamiltonian {\em directly} with suitable hardware; however in many applications, including chemistry-related tasks, $H_T$ involves non-local connectivity and high-degree terms that are infeasible to implement.

\begin{figure*}[t]
	\begin{centering}
		\includegraphics[width=1.5\columnwidth]{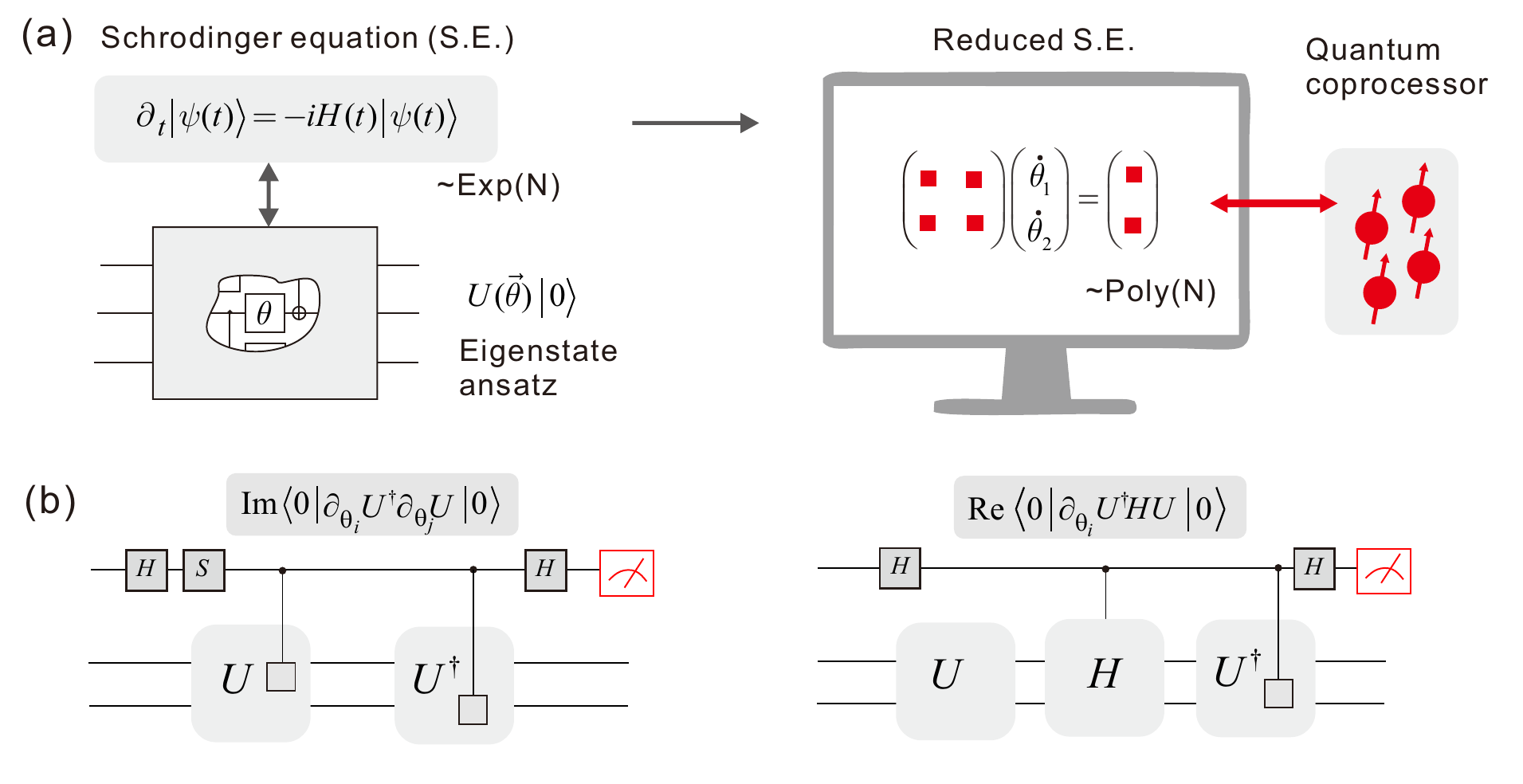}\\
		\caption{\label{fig1} Variational quantum simulation of dynamics. (a) Sketch of the variational algorithm. A shallow circuit ansatz $\ket{\phi(\vec{\theta})}=U(\vec{\theta)}\ket{0}$ is used to approximate the time evolved state $\ket{\psi(t)}$ with only a polynomial number of parameters $\vec{\theta}$. The evolution of the Schr\"odinger equation is reduced to the evolution of the parameters according to $\frac{\partial}{\partial t}\vec{\theta}(t)=M^{-1}(t) \cdot V(t)$. Here the matrix $M(t)$ and vector $V(t)$ elements are defined by 
		$M_{i,j}(t)=-\textrm{Im}\left(\frac{\partial\bra{\phi(\vec{\theta})}}{\partial\theta_i} \frac{\partial\ket{\phi(\vec{\theta})}}{\partial \theta_j}\right)=-\textrm{Im}\braket{0|\partial_{\theta_i} U^\dag\partial_{\theta_j}U|0}$ and
        $V_i(t)=\textrm{Re}\left(\frac{\partial\bra{\phi(\vec{\theta})}}{\partial\theta_i} H \ket{\phi(\vec{\theta})}\right)=\textrm{Re}\braket{0|\partial_{\theta_i} U^\dag HU|0}$.
        As both $M$ and $V$ are in the form of the real or imaginary part of $\braket{0|V|0}$, they can be efficiently evaluated with the (b) Hadamard test. Suppose $U(\vec \theta)=U_{L}(\theta_L)\dots U_{i}(\theta_i)\dots U_{1}(\theta_1)$, then the partial derivative of parameter $i$ is $\partial U_i/\partial \theta_i=\sum_{k} f_{i,q} U_{i}R_{i,q}$ with unitary $R_{i,k}$, thus we have $\partial_{\theta_i}U=\sum_{i}f_{i,k} U_{L}(\theta_L)\dots U_{i}(\theta_i)R_{i,k}\dots U_{1}(\theta_1)$, which is a linear sum of unitary operators with one extra gate inserted in the original unitary $U(\vec \theta)$. Denote $U_{i,k} = U_{L}(\theta_L)\dots U_{i}(\theta_i)R_{i,k}\dots U_{1}(\theta_1)$, then we have $M_{i,j}=-\sum_{k,q}\textrm{Im}\left(f_{i,k}^*f_{j,q}\braket{0|U_{i,k}^\dag U_{j,k}|0}\right)$ and $V_{i}=\sum_{k}\textrm{Re}\left(f_{i,k}^*\braket{0|U_{i,k}^\dag HU|0}\right)$. Each term in the sum can be efficiently evaluated with shallow quantum circuits as shown in (b).
% 		These circuits can be efficiently evaluated on near-term quantum computers with shallow quantum circuits.
		}
	\end{centering}
\end{figure*}

 We might resort instead to the flexibility of a fully digitised gate-based quantum circuit. In this context the recent quantum-classical hybrid algorithms, such as  quantum approximate optimisation algorithm (QAOA)~\cite{farhi2014quantum} and the variational quantum eigensolver (VQE)~\cite{peruzzo2014variational}, 
%  ~\cite{peruzzo2014variational,wang2015quantum,PRXH2,PhysRevA.95.020501,VQETheoryNJP,PhysRevLett.118.100503,PhysRevX.8.011021,Santagatieaap9646,kandala2017hardware,kandala2018extending,PhysRevX.8.031022,kokail2018self}, 
are a promising route toward useful exploitation of small- and medium-scale quantum computers. Generally, a hybrid algorithm would encode a `trial' quantum state via a shallow parameterised  quantum circuit (i.e. the quantum coprocessor). A governing classical computer iteratively adjusts the parameters and monitors the output of the quantum circuit, ultimately seeking the parameters for which the output matches the ground state of $H_T$. The challenge is to achieve this in a fashion that can scale to the case of hundreds or thousands of parameters; the feasibility of this task is an active area of study~\cite{mcardle2018quantum}. 
 
 Here, we marry together the adiabatic protocol with a circuit-based NISQ coprocessor. This is enabled by recent theoretical work showing that the general dynamical evolution of (both closed and open) physical systems can be efficiently simulated with variational quantum algorithms~\cite{Li2017,2018arXiv181208767Y}. As the classical computer iteratively updates the coprocessor's parameters we should find its evolving output tracks the state that {\em would} be realised in ideal adiabatic machine. Thus the adiabatic system is modelled by a large number of short executions on our circuit-based quantum coprocessor.

Our Letter presents the first experimental validation of the theory of variational quantum dynamics~\cite{Li2017,2018arXiv181208767Y} in any context. We use our system to model a 1D Ising spin chain undergoing an adiabatic phase transition. For a two-qubit Ising model, we simulate the evolution of the full energy spectrum by adiabatically changing the Hamiltonian. For a three-qubit Ising model, we show the evolution of the ground state and observe phase transition from paramagnetic to ferromagnetic.

In variational quantum dynamics simulation as shown in Fig.~\ref{fig1}(a), we translate the problem of simulating the Schr\"odinger equation of a pure state
\begin{equation}
\begin{aligned}
\frac{\partial}{\partial t}|\psi(t)\rangle=-iH|\psi(t)\rangle (\hbar=1),
\end{aligned}
\end{equation}
into optimisation of a parameterised state, $\ket{\phi(\vec{\theta})}=U(\vec{\theta})\ket{0}$, such as to have $|\phi(t)\rangle$ a good approximation of  $|\psi(t)\rangle$.
Here $U$ is described by $L$ variable single-qubit rotation gates of angles $\vec{\theta}=(\theta_1,\theta_2,...,\theta_L)$, where the number of parameters $L\ll2^N$. Based on the algorithm, the derivative of the parameter space can be expressed by $\frac{\partial}{\partial t}\vec{\theta}(t)=M^{-1}(t) \cdot V(t)$, where the elements of matrix $M$ and vector $V$ are respectively,
\begin{equation}
\begin{aligned}
M_{i,j}(t)&=-\textrm{Im}\left(\frac{\partial\bra{\phi(\vec{\theta})}}{\partial\theta_i} \frac{\partial\ket{\phi(\vec{\theta})}}{\partial \theta_j}\right),\\
V_i(t)&=\textrm{Re}\left(\frac{\partial\bra{\phi(\vec{\theta})}}{\partial\theta_i} H \ket{\phi(\vec{\theta})}\right).
\end{aligned}
\end{equation}
The values of these elements can be efficiently evaluated by the Hadamard test circuits, as shown in Fig.~\ref{fig1}(b). Thus the evolution of $\vec{\theta}$ can be found over time, given $\vec\theta(t+\delta t)=\vec\theta(t)+\delta t\dot{\vec\theta}(t)$, if $\delta t$ is sufficiently small.

% Thus, using this time-dependent quantum state  $\ket{\psi}=\ket{\vec{\theta}(t)}$ and a corresponding reduced version of Schr\"odinger equations $\frac{\partial}{\partial t}\vec{\theta}(t)=\Lambda(t)$, we could efficiently simulate the adiabatic dynamics on classical computers. The basic concept is .

% Recently, the work by Li and Benjamin shows that for a system Hamiltonian $H(t)$ and a parameterized quantum state $\ket{\psi(\vec{\theta}(t))}$, 

\begin{figure*}[t]
	\begin{centering}
		\includegraphics[width=1.6\columnwidth]{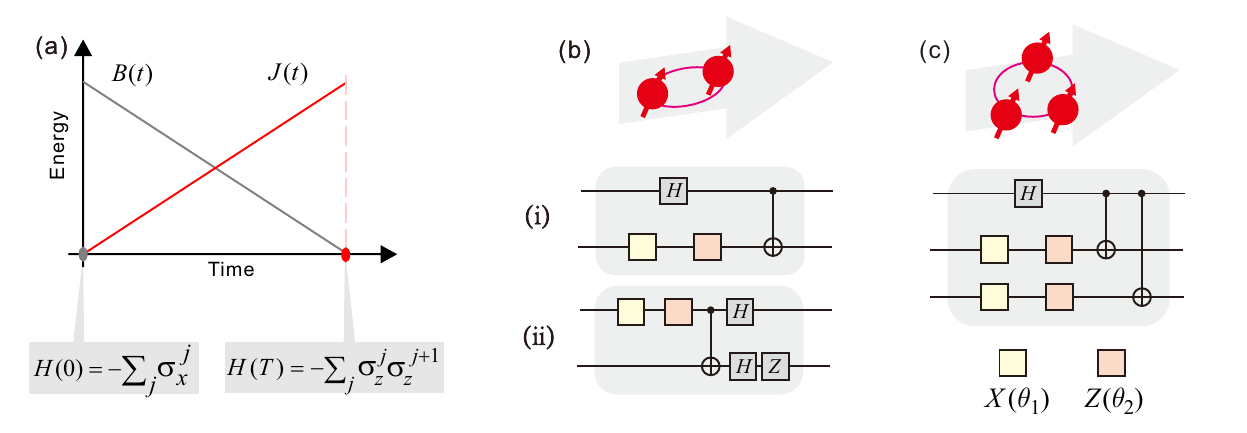}\\
		\caption{\label{fig2} The Ising model simulated in the experiment. (a) The adiabatic change of system Hamiltonians. (b) The circuit ansatz for the eigenspectrum of the two-spin system. (i) is used for preparing the ground and the third-excited states, and (ii) is used to prepare for the first- and second-excited states.(c) The circuit ansatz for the ground-state problem of the fully coupled 3-spin system.
		}
	\end{centering}
\end{figure*}

In our experiment, we focus on the 1D Ising model under a transverse magnetic field. The time-dependent Hamiltonian with a periodic boundary condition is 
\begin{equation}
    H(t) = B(t)H_0 + J(t)H_T,
\end{equation}
with $H_0 = -\sum_j \sigma_x^j$ and $H_T = -\sum_j \sigma_z^j\sigma_z^{j+1}$. 
Under the condition of $|J|\ll |B|$, the ground state of the system is that all spins are aligned along the $x$ axis $\ket{++\cdots+}$, corresponding to the paramagnetic ordered state. On the other hand when $|J|\gg |B|>0$, the system's ground state approaches degeneracy and is entangled as $(\ket{00\cdots0}+\ket{11\cdots1})/\sqrt{2}$, representing the ferromagnetic ordered state 
% {\color{red}comment, no need to mention in paper: for strictly B=0, it doesn't HAVE to be entangled since $|000>$ and $|111>$ are also good states spanning the g.s. space}. 
Therefore, by preparing the ground state $\ket{++\cdots+}$ of $H_0$ at time $t=0$ and adiabatically evolving it from $H_0$ to $H_T$ with $H(t)$, we can observe a transition from paramagnetic to ferromagnetic. To do so, we set  $J(t)=t/T$ and $B(t)=1-t/T$, which represent the strength of spin-spin interactions and  the strength of the magnetic field, respectively. The change of the Hamilitonian over the whole time period is shown in Fig.~\ref{fig2}(a).

\begin{figure*}[t]
	\begin{centering}
		\includegraphics[width=2\columnwidth]{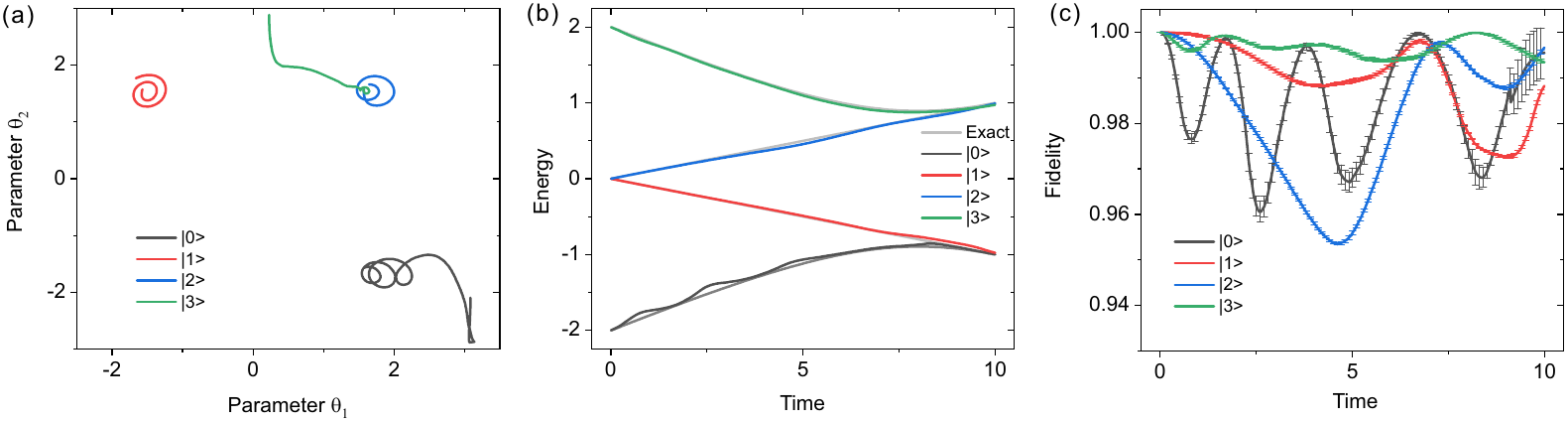}\\
		\caption{\label{fig3}Experimental results of the eigenspectrum of two spins. (a) The evolution of circuit parameters. The chosen first- and second-excited states are special thus they should remain unchanged over the whole period of time. This is reflected by the circled shape of the parameter paths for the two cases. (b) The time-dependent energy level diagram. The grey curve represents the exact evolution paths of the energy over time, while the coloured curves are obtained from experimental measurements. (c) The fidelities of the experimentally obtained quantum states to the exact states, i.e.,  $|\langle\psi(t)|\phi(t)\rangle|^2$. The error bar is produced from 100 Monte Carlo simulations of the statistical-fluctuation from the finite measurement (See Methods for more details).
		}
	\end{centering}
\end{figure*}

\begin{figure*}[htbp]
\begin{centering}
\includegraphics[width=2\columnwidth]{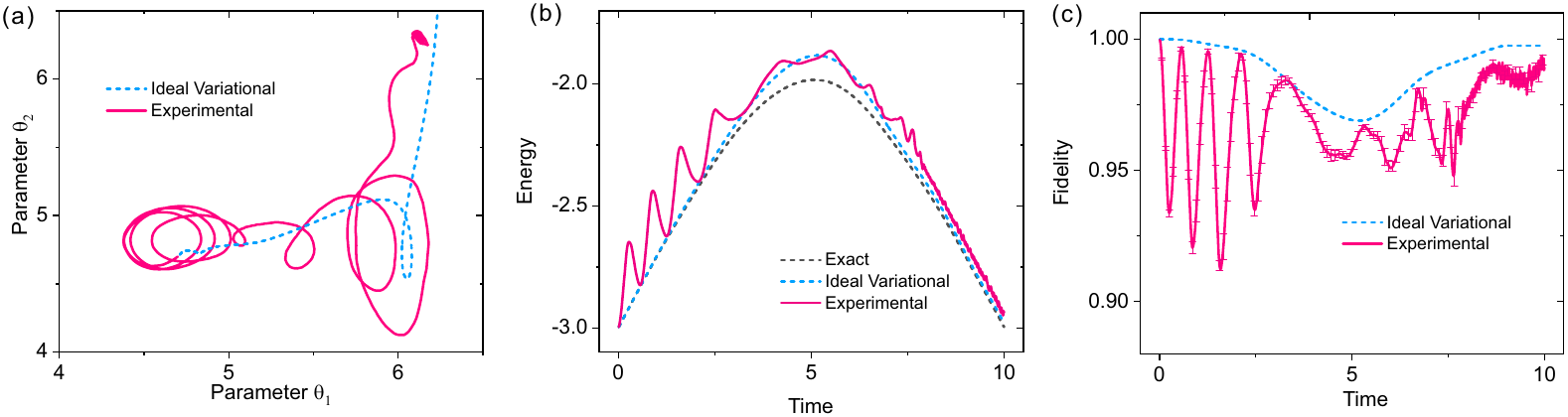}\\
\caption{\label{fig4}Experimental results of the ground state of 3 spins. (a) The parameter trajectory on the phase diagram. The blue dashed curve is the result of  modelling of the experimental process with the same initial parameters and circuit sets. The red curve represents the experimental data, which follows roughly the same trend of the blue curve. (b) The evolution of the ground-state energy. The red curve refers to the ideal ground state of the system Hamiltonian. The blue dashed curve results from theoretical modelling and the blue curve represents the experimental result. (c) The fidelity of the experimentally obtained quantum state to the exact state. With oscillations, the red curve follows the outline of the the blue dashed curve, which is obtained from theoretical modelling. Interestingly, even though the ansatz cannot exactly represent the target state for intermediate time (note the dip at $t\sim 5$), the evolution projected onto the ansatz manifold nevertheless successfully drives the state to the ground state of the final Hamiltonian.
 Our experiment results in a high final state fidelity of $98.9\%$.} 
\end{centering}
\end{figure*}

We first consider the Ising model of two spins. In various experiments we start from the ground, first, second and third eigenstates of the Hamiltonian $H_0$, given by $\ket{++}$, $(\ket{+-}\pm\ket{-+})/\sqrt{2}$ and $\ket{--}$, respectively. The circuit ansatze are chosen such that they can exactly cover the adiabatic paths, as shown in Fig.~\ref{fig2}(b). One ansatz is used for preparing the ground and the third-excited states, and the other is used to initialise the qubits to be at the first- and second-excited states of $H_0$. In both cases, two parameters $\theta_1$ and $\theta_2$ are applied in the circuits, controlling two single-qubit gates with a rotation angle around the $X$ and $Z$ axis, respectively. 

The experiment was conducted in an Xmon-superconducting quantum processor, which is illustrated in Fig.~\ref{fig_s1} and described in detail in the Methods. In the experiment, the coefficients $M_{ij}$ and $V_i$ were generated from a set of auxiliary quantum circuits, which can be found in Fig.~\ref{fig_s2} and Fig.~\ref{fig_s3}, corresponding to the two circuit ansatze shown in Fig.~\ref{fig2} respectively.
% The auxiliary circuits are performed on a superconducting quantum processor,   and scan the two-dimension parameter space on a $20 \times 20$ grid. 
The parameter trajectories and the evolution of the energy eigenvalues are shown in Fig.~\ref{fig3}(a) and \ref{fig3}(b), respectively.
The experimental results agree well with the system eigenspectrum at all times, which verifies the effectiveness of the adiabatic variational quantum method. We also show the quantum state fidelity in Fig.~\ref{fig3}(c) and find that the average fidelities are all above 95\% with systematic oscillations. 
%\textcolor{red}{Fidelity of the states at time $T$.}\\
In particular, the fidelities of the four eigenstates at time $T$ are $99.5\%$, 
$98.8\%$,
$99.6\%$, and $99.3\%$, respectively

Next, we demonstrate the evolution of the ground state of a three-spin Ising Hamiltonian. The ground states of $H_0$ and $H_T$ are $\ket{+++}$ and $(\ket{000}+\ket{111})/\sqrt{2}$, corresponding to paramagnetic and ferrimagnetic, respectively.
The circuit ansatz to simulate the adiabatic phase transition process is shown in Fig.~\ref{fig2}(c). It accurately expresses the ground states of the initial Hamiltonian $H_0$ and the target Hamiltonian $H_T$, and approximates the ground state of the intermediate Hamiltonian. The exact circuits evaluating $M_{ij}$ and $V_i$ in the experiment are shown in Fig.~\ref{fig_s4}, where the coefficients $M_{ij}$ were evaluated by Hadamard test,  while $V_i$ were obtained from direct measurement due to experimental limitations. 

In Fig.~\ref{fig4}(a), we show the phase diagram of parameters. The trajectory tracks the theoretical one but with additional oscillations; these mainly originate from residual phase crosstalk in the experimental multi-qubit circuits and do not affect the average trajectory. The energy and state fidelity of the evolving state are shown in Fig.~\ref{fig4}(b) and Fig.~\ref{fig4}(c), respectively.  In experiment, we observe a fidelity of $98.9\%$ for the state at time $T$.

%\textcolor{red}{Add something in between?}\\
We emphasise that although the simulation proceeds by a series of parameter updates governed by the classical computer, the classical machine does not model the quantum state. The computational advantage originates from using the quantum coprocessor to efficiently estimate the coefficients of the evolution equation, which will be beyond the simulation capability of state-of-the-art supercomputers at the scale of about $50$-qubit circuits of depth $50$ gate layers.

When seeking to scale the algorithm to a large quantum system, we would find that the fidelity of simulation depends on two related issues: the choice of an appropriate circuit ansatz and how to resist experimental noise. Encouraging progress is being made in both topics. 
 A good ansatz circuit would be informed by knowledge of the physical system being simulated; for example, the circuit could be made to respect conserved symmetries~\cite{lee2018generalized}. Moreover, one could take inspiration from work on variational ansatze for finding ground states of quantum systems, which have been shown to benefit from incorporating physical intuition~\cite{mcclean2018barren,barkoutsos2018quantum}. 
 Meanwhile noise can be dramatically suppressed with recently proposed error mitigation methods such as techniques based on extrapolating different experimental data~\cite{subspace1,Li2017,PhysRevLett.119.180509, endo2017practical, subspace2,recoveringnoisefree, samerrormitigation,bonet2018low}. We note that in our small-scale demonstration, we have used interpolation of the sampled parameters to reduce the discretisation error.

In summary, we have reported the first experimental demonstration of a variational quantum algorithm to simulate a system's dynamics. As the context of our simulation, we chose the task of adiabatic state preparation --- an important component of many quantum applications. In our hybrid approach the linear equations set is tractable (and scalable) on a classical computer while the intractable part is offloaded to a quantum coprocessor. This method can be successful even where alternatives such as analog or digital adiabatic state preparation would be infeasible; the former may fail due to unfavourable properties of the Hamiltonian (high connectivity or high degree terms), while the latter may require deep circuits for accurate results. Our work therefore represents a new and promising approach to quantum dynamics simulation and quantum enhanced optimisation on intermediate-scale quantum computers.\\

\section{Methods}
The experiment was conducted on a 12-qubit superconducting quantum processor. The processor is in the Xmon architecture, which is a variety of grounded transmon with tunable qubit frequency. The processor diagram is shown in Fig.~\ref{fig_s1}. The average energy relaxation time $T_1$ and dephasing time $T_2^{\star}$ were 36.9 $\mu s$ and 4.1 $\mu s$ at working points of qubits, respectively. The average gate fidelity of single-qubit gates and two-qubit controlled phase (CZ) gates were 99.8\% and 99\% from randomized benchmarking, respectively. The average readout fidelities of states $\ket{0}$ and $\ket{1}$ were 93.7\% and 84.1\%, respectively. The readout separating error were further corrected by the inverse POVM matrices to generate a corrected distribution. We used up to 4 qubits in the experiments. 

The circuit ansatz of two Ising spins was constructed from Schmidt decomposition of ideal states to cover the quantum state manifold of adiabatic path. The circuit ansatz of three Ising spins was choose to not cover the state space of the intermediate adiabatic path to investigate the interplay of approximated ansatz and faithful adiabatic following. 

The linear equations set $\frac{\partial}{\partial t}\vec{\theta}(t)=M^{-1}(t) \cdot V(t)$ was solved with a classical computer, and the elements of $M$ and $V$ were obtained by evaluating a set of auxiliary circuits realised on quantum processor. For the two Ising spin case, the elements of $M$ and $V$ were both evaluated with the Hadamard test circuits. For the three Ising spin case, the elements of $M$ were generated from a set of auxiliary Hadamard test circuits, while the elements of $V$ were obtained by taking the derivatives of the measured expectation values of the Hamiltonian over each small time period. All the Hadamard test circuits were compiled by moving, commuting, and eliminating some of the basic gate operations. The original and complied circuits are shown in Fig.~\ref{fig_s2} to~\ref{fig_s4}. 

The gradient $\{\dot{\theta_1},\dot{\theta_2}\}$ was evaluated at each time steps to update the parameters $\{\theta_1,\theta_2\}$. The direct iterative method was set to have a fixed time step of 0.01, which is sufficiently small to eliminate the discretisation induced error over a total evolution period of 10. At each time instant, the auxiliary circuits were first measured on a discrete 20 x 20 grid of the parameter space i.e. $\{\theta_1,\theta_2\}$ was scanned from 0 to 2$\pi$ with a step size $\pi/10$  and then interpolated to arbitrary ones to generate the coefficients over the whole parameter space (an example of the parameter space of elements of $M$ and $V$ at $t=5$ are shown in Fig.~\ref{fig_s5} to~\ref{fig_s7}). Thus a database was constructed and used when solving the linear equations. 

The error bars in Fig.\ref{fig3}(c) and Fig.\ref{fig4}(c) were generated based on the bootstrap method, which is a commonly used method to estimate error propagation in complex functions. Suppose $y=f(x)$, where $x$ here is the measured elements in $M$ and $V$, and $f(x)$ is a map of $x$ to the fidelity of the trial state. For each data point, the error range was calculated by 100 samples of the measured values of $M$ and $V$, i.e., $x_0$, from Gaussian distribution (average=$x_0$, std=$1/\sqrt{100}$).

~\\
~\\
\textbf{Acknowledgements.}
We thank Ying Li for insightful discussion of the theory.
This work was supported by the National Natural Science
Foundation of China, the Chinese Academy of Sciences, the National Fundamental Research Program and the Anhui Initiative in Quantum Information Technologies. SCB and XX are supported by the Office of the Director of National Intelligence (ODNI), Intelligence Advanced Research Projects Activity (IARPA), via the U.S. Army Research Office Grant No. W911NF-16-1-0070. The views and conclusions contained herein are those of the authors and should not be interpreted as necessarily representing the official policies or endorsements, either expressed or implied, of the ODNI, IARPA, or the U.S. Government. The U.S. Government is authorized to reproduce and distribute reprints for Governmental purposes notwithstanding any copyright annotation thereon. Any opinions, findings, and conclusions or recommendations expressed in this material are those of the author(s) and do not necessarily reflect the view of the U.S. Army Research Office. SCB and XY acknowledge EPSRC grant EP/M013243/1.

\bibliographystyle{naturemag}
\bibliography{VariationalExp}

\clearpage
\appendix
\widetext

\section{Supplementary Information}

\begin{figure}[h]
	\centering
	\includegraphics[width=0.5\textwidth]{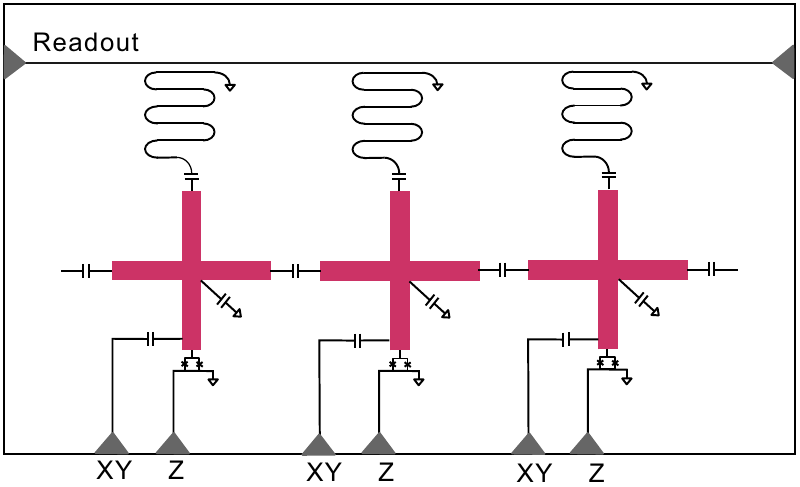}
	\renewcommand\thefigure{S1} 
	\caption{The circuit diagram of Xmon architecture processor. The qubits are variety of grounded transmon with tunable qubit frequency. The neighboring qubits are coupled capacitively to implement fast adiabatic CZ gates. A microwave drive line (XY) and a fast flux-bias line (Z) are connected to each qubits for gate operations. A readout resonator is dispersively coupled to each qubit and probed by a common transmission line.}
	\label{fig_s1}	
\end{figure}

\begin{figure*}[]
	\centering
	\includegraphics[width=0.8\textwidth]{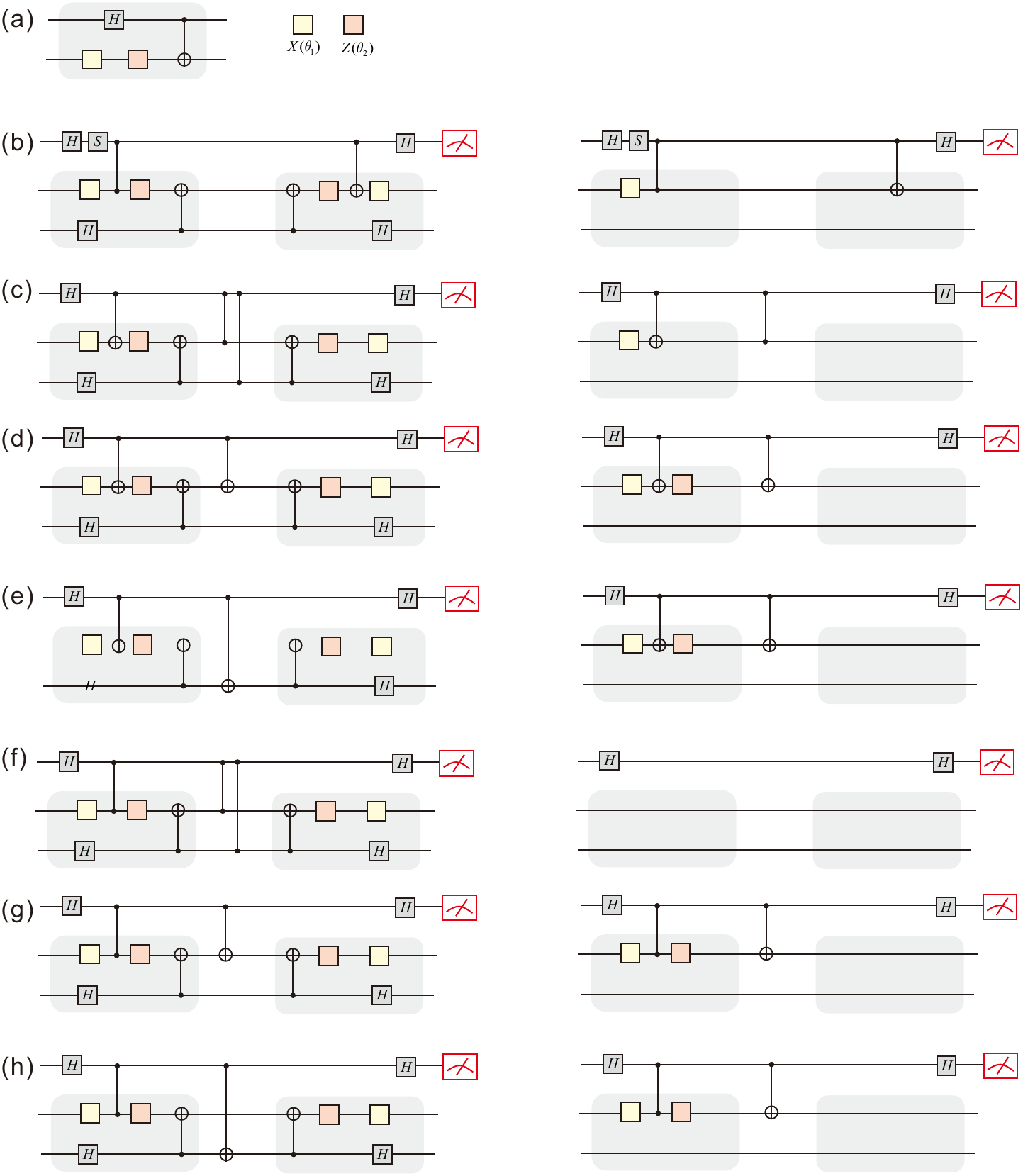}
	\renewcommand\thefigure{S2} 
	\caption{Exact circuits for the two-spin experiment for the $\ket{0}$ and $\ket{3}$ states. All the circuits are compiled by moving, commuting, and eliminating some of the basic gate operations to reduce the number of gates. The circuits before and after the compiling process are shown respectively in the left and right. (a) The circuit ansatz. (b) The circuit for $M_{12}$. (c)(d)(e) The circuit for $V_1$. (f)(g)(h) The circuit for $V_2$. }
	\label{fig_s2}	
\end{figure*}

\begin{figure*}[]
	\centering
	\includegraphics[width=0.8\textwidth]{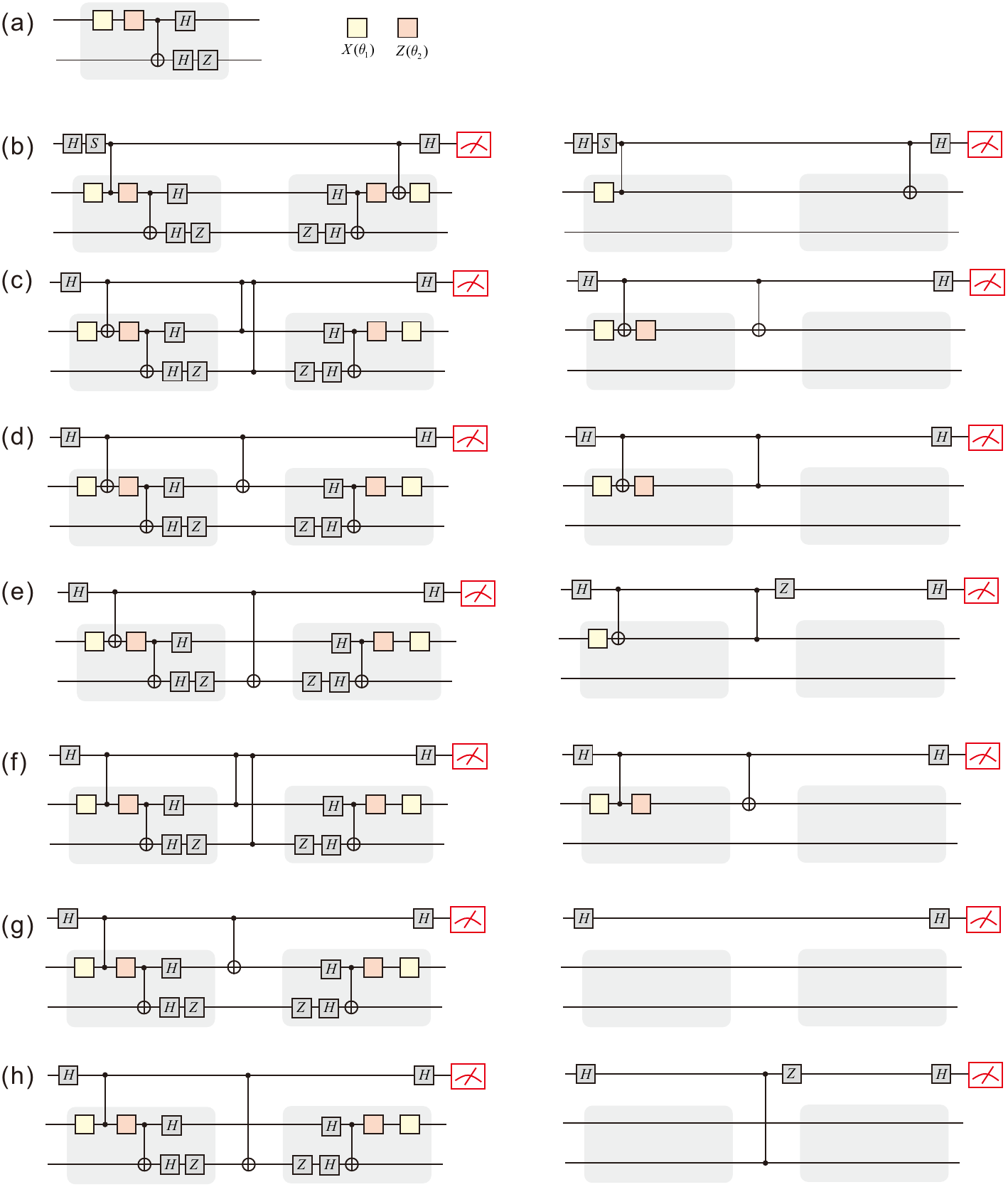}
	\renewcommand\thefigure{S3} 
	\caption{Exact circuits for the two-spin experiment for the $\ket{1}$ and $\ket{2}$ states. The original circuits(left) are compiled as described in Fig.~\ref{fig_s2} and are shown in the right. (a) The ansatz. (b) The circuit for $M_{12}$. (c)(d)(e) The circuit for $V_1$. (f)(g)(h) The circuit for $V_2$.}
	\label{fig_s3}	
\end{figure*}

\begin{figure*}[]
	\centering
	\includegraphics[width=0.8\textwidth]{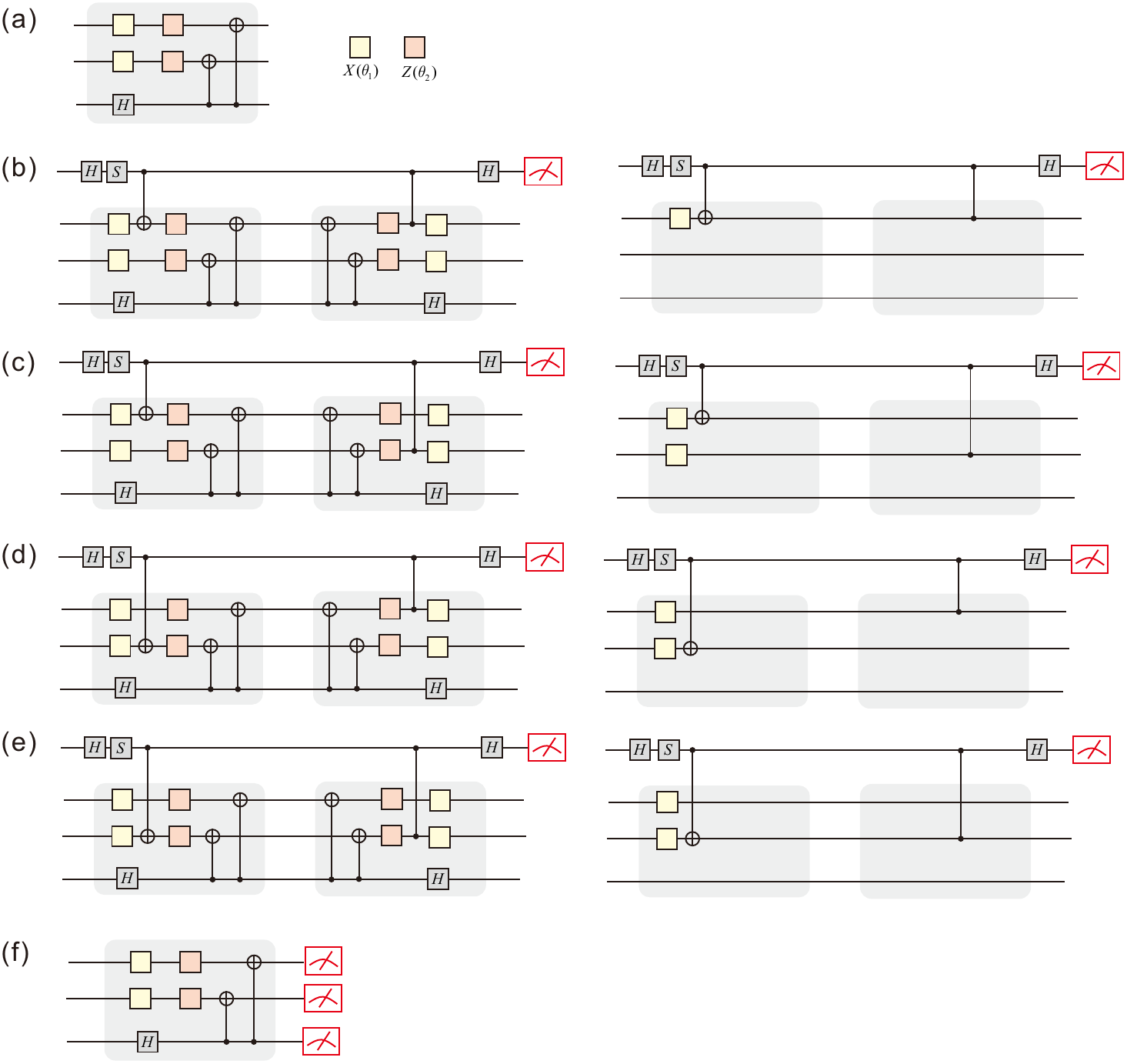}
	\renewcommand\thefigure{S4} 
	\caption{Exact circuits for the three-spin experiment for the ground state. The original circuits(left) are compiled as described in Fig.~\ref{fig_s2} and are shown in the right. (a) The ansatz. (b)(c)(d)(e) The circuit for $M_{12}$. (f) The circuit for $V_1$ and $V_2$.}
	\label{fig_s4}	
\end{figure*}

\begin{figure*}[t]
	\centering
	\includegraphics[width=0.9\textwidth]{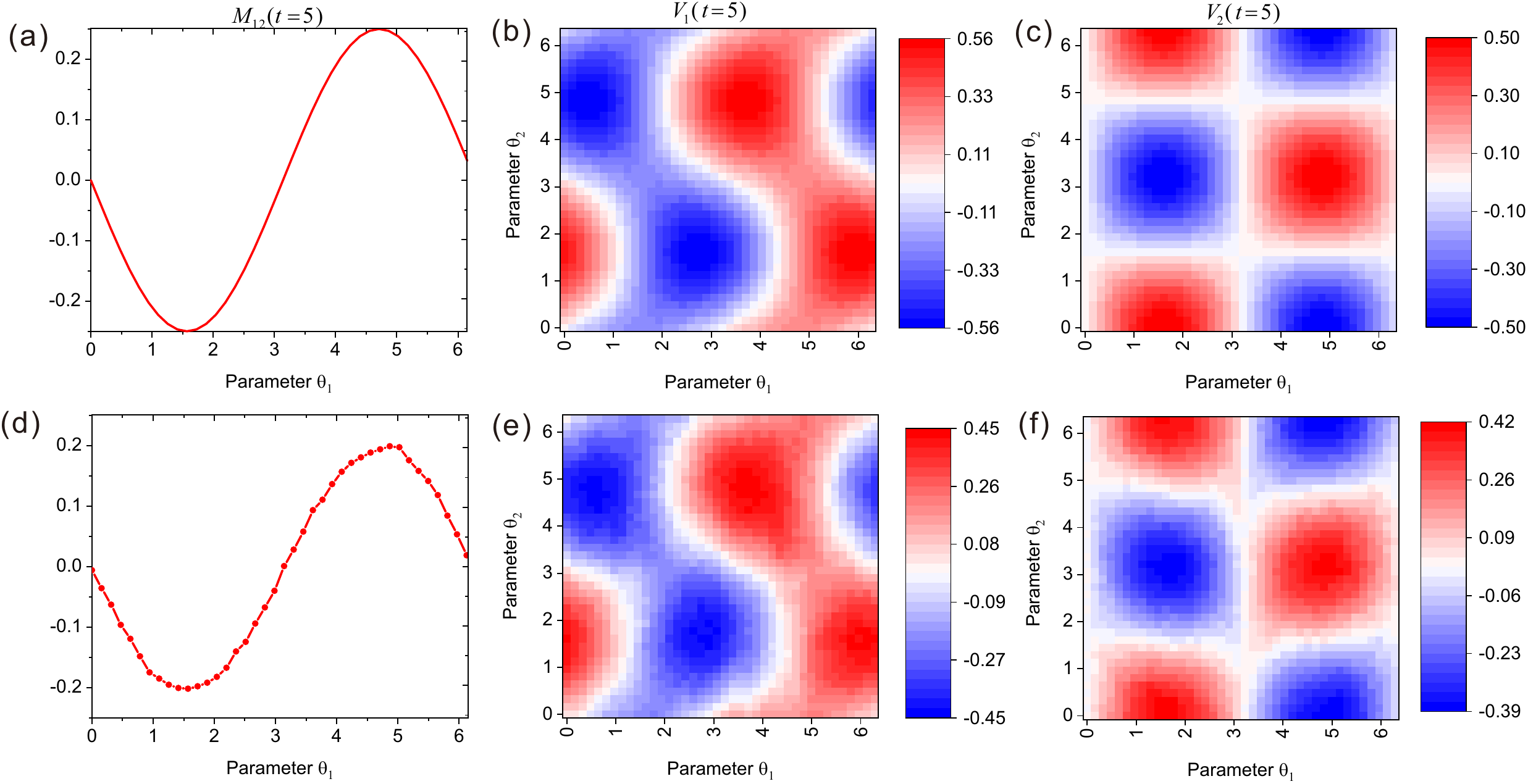}
	\renewcommand\thefigure{S5} 
	\caption{The parameter space for elements of $M$ and $V$ at $t=5$. The circuits correspond to the $\ket{0}$ and $\ket{3}$ states of two Ising spins. (a) The absolute value of the theoretically modelled $M_{1,2}$ as a change of $\theta_1$. (b)(c) The change of the absolute value of the theoretically modelled $V_1$ and $V_2$ respectively in the parameter grid made up by $\theta_1$ and $\theta_2$. (d) Experimentally obtained $M_{1,2}$ as a change of $\theta_1$. (e)(f) Experimentally obtained $V_1$ and $V_2$ in the parameter space of $\theta_1$ and $\theta_2$.}
	\label{fig_s5}	
\end{figure*}

\begin{figure*}[]
	\centering
	\includegraphics[width=0.9\textwidth]{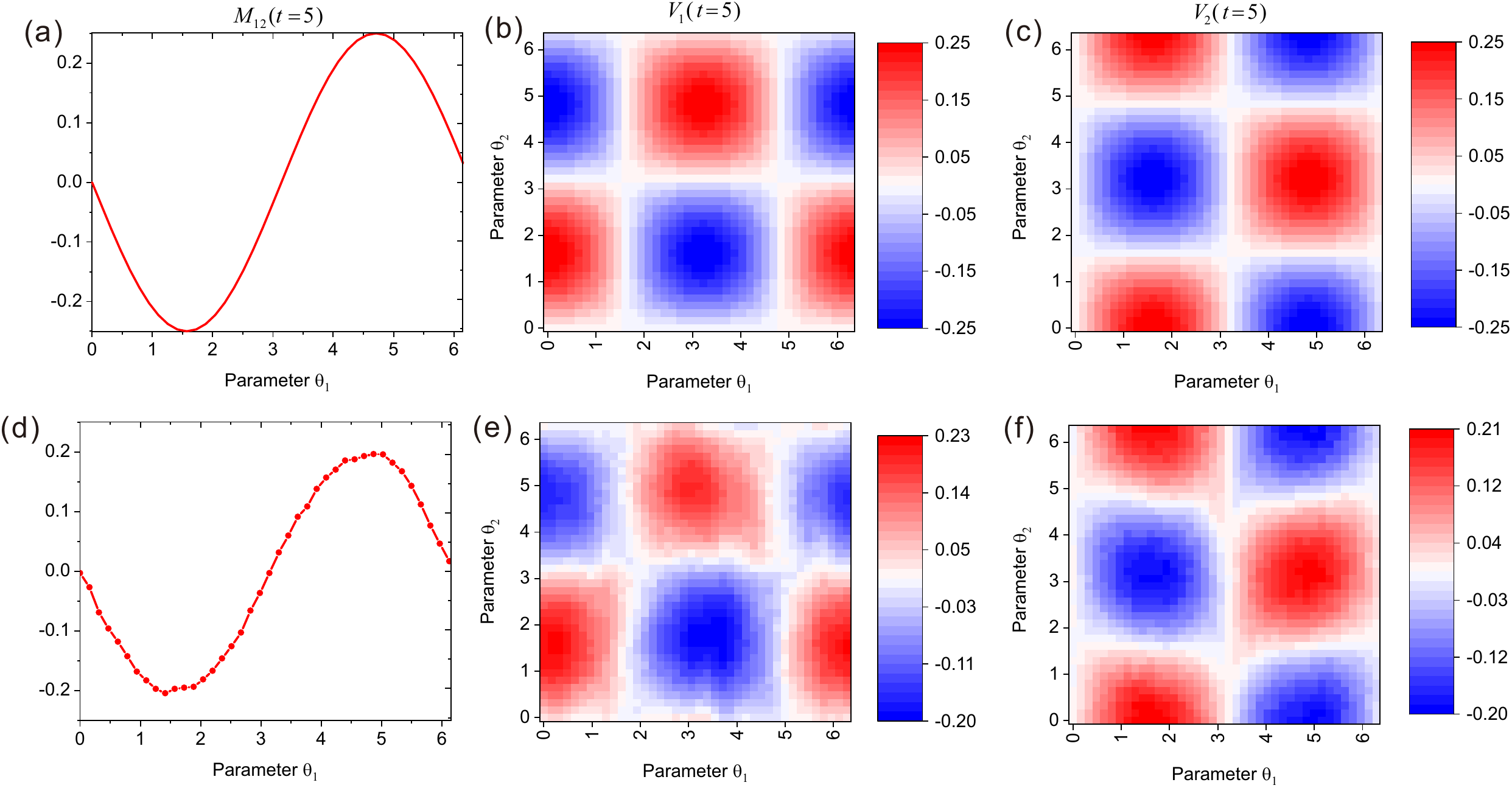}
	\renewcommand\thefigure{S6} 
	\caption{Same as Fig.~\ref{fig_s5}, but the circuits correspond to the  $\ket{1}$ and $\ket{2}$ states of two Ising spins.}
	\label{fig_s6}	
\end{figure*}

\begin{figure*}[]
	\centering
	\includegraphics[width=0.9\textwidth]{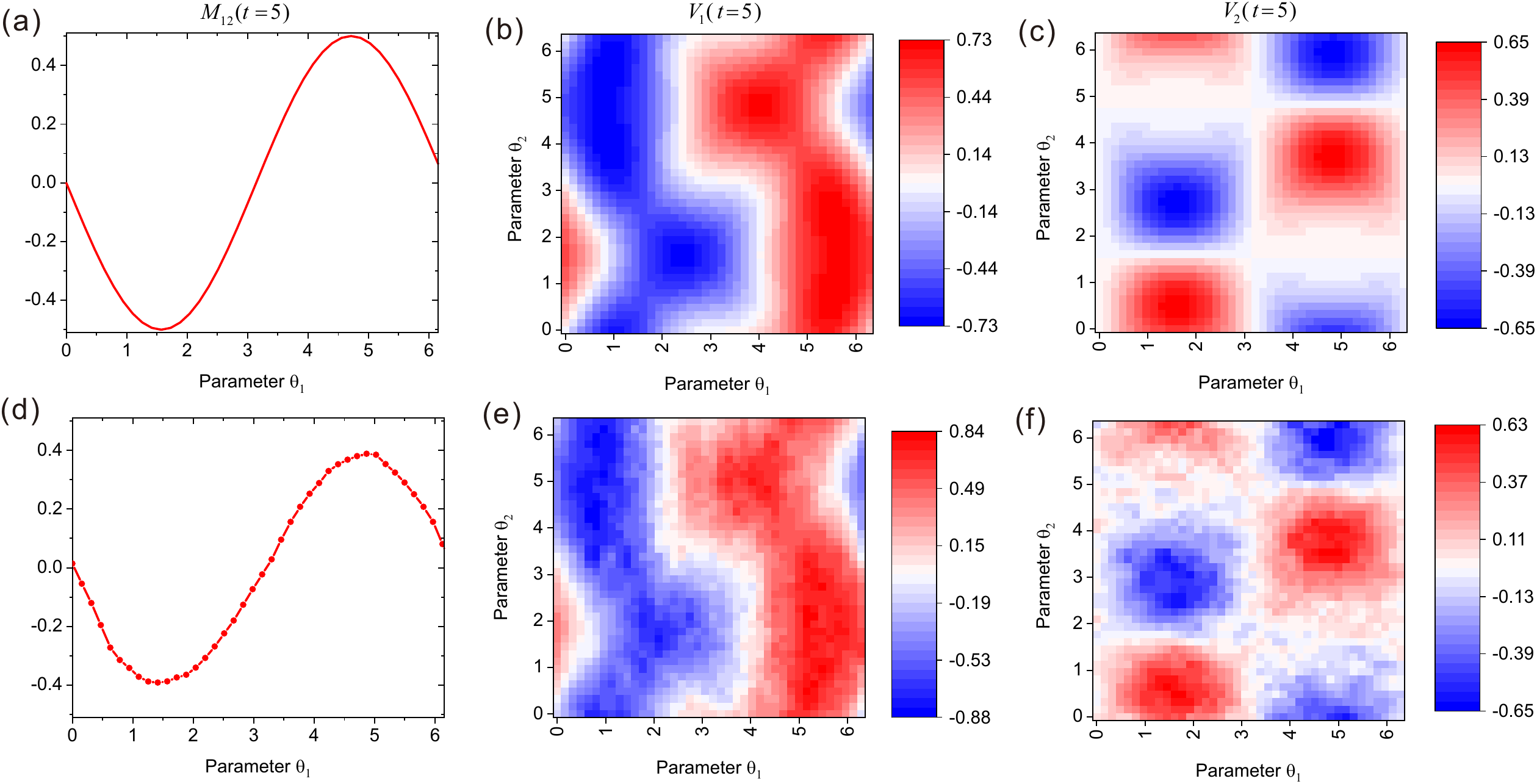}
	\renewcommand\thefigure{S7} 
	\caption{Same as Fig.~\ref{fig_s5}, but the circuits correspond to the  ground state of three Ising spins.}
	\label{fig_s7}	
\end{figure*}

\end{document}